\documentclass[conference,10pt]{IEEEtran}

\usepackage{amsmath,graphicx,subfigure,epsfig,amssymb,amsmath,epstopdf,tikz,float,tikz-3dplot,color,pdfpages,cite,booktabs,psfrag,multicol}
\usepackage{import}
\usepackage{dsfont}
\usepackage{url}	
\usepackage{algorithmic}
\usepackage{algorithm}
\usepackage{multirow}
\usepackage{graphicx}
\usepackage{geometry}
\usepackage[T1]{fontenc}
\usepackage{xcolor}
\usepackage{comment}
\usepackage{fancyhdr}
\usepackage{textcomp}
\usetikzlibrary{decorations.pathreplacing,decorations.pathmorphing,decorations.markings,shapes,backgrounds,patterns,decorations.text}

\title{{\fontsize{8pt}{8pt}\selectfont\textcolor{blue}{This is a preprint version. The published version will be available in the proceedings of IEEE FNWF 2025 on IEEEXplore subsequently.}} \\
A Novel Integrated Architecture for Intent Based Approach and Zero Touch Networks}

\author{Neelam Gupta$^{1}$, Dibakar Das$^{1}$, Tamizhelakkiya K$^{1}$,Uma Maheswari Natarajan$^{1}$, Sharvari Ravindran$^{1}$,  Komal Sharma$^{2}$, \\Jyotsna Bapat$^{1}$, and Debabrata Das$^{1}$\\
$^{1}$Networking and Communication Research Lab,  IIIT Bangalore, India\\
$^{2}$Toshiba Software (India) Private Limited, Bangalore, India\\
Email: \{neelam.gupta, dibakar.das, tamizhelakkiya.k, umamaheswari.natarajan, sharvari.r, jbapat, ddas\}@iiitb.ac.in, \\komal.sharma@toshiba-tsip.com}

\providecommand{\keywords}[1]{\textbf{\textit{Keywords---}} #1}
\begin{document}
%\ninept

\maketitle
%\thispagestyle{empty}
%\pagestyle{headings}

%The repository is a result of collaborative effort by multiple organizations
\begin{abstract}
The transition to Sixth Generation (6G) networks presents challenges in managing quality of service (QoS) of diverse applications and achieving Service Level Agreements (SLAs) under varying network conditions. Hence, network management must be automated with the help of Machine Learning (ML) and Artificial Intelligence (AI) to achieve real-time requirements. Zero touch network (ZTN) is one of the frameworks to automate network management with mechanisms such as closed loop control to ensure that the goals are met perpetually.
%since it is a difficult task to achieve conflicting requirements like high speed, low latency, massive connections, high reliability, and mobility, utilizing existing network resources.
Intent-Based Networking (IBN) specifies the user intents with diverse network requirements or goals which are then translated into specific network configurations and actions. %, IBN enables management and provisioning of future network architectures.
This paper presents a novel architecture for integrating IBN and ZTN to serve the intent goals. %using a closed-loop control system and Long Short-Term Memory (LSTM)-based bandwidth prediction.
%ZTN's predictive functions use the Network Intent Language (Nile) language to express bandwidth requirements with IBN, acting as runtime thresholds. The approach focuses on four key aspects: analysis of anticipated throughput over grouped time intervals, bandwidth intent extraction, estimation of the rate of reaching thresholds for each episode, and evaluation of optimal and suboptimal actions in different scenarios.
Users provides the intent in the form of natural language, e.g., English, which is then translated using natural language processing (NLP) techniques (e.g., retrieval augmented generation (RAG)) into \textit{Network Intent LanguagE (Nile)}. The Nile intent is then passed on to the BiLSTM and Q-learning based ZTN closed loop framework as a goal which maintains the intent under varying network conditions. Thus, the proposed architecture can work autonomously to ensure the network performance goal is met by just specifying the user intent in English.
The integrated architecture is also implemented on a testbed using \textit{OpenAirInterface (OAI)}. Additionally, to evaluate the architecture, an optimization problem is formulated which evaluated with Monte Carlo simulations.
Results demonstrate how ZTN can help achieve the bandwidth goals autonomously set by user intent. The simulation and the testbed results are compared and they show similar trend. Mean Opinion Score (MOS) for Quality of Experience (QoE) is also measured to indicate the user satisfaction of the intent. %The performance of the testbed and the Monte Carlo simulation are compared to show how accurately the proposed method meets the bandwidth goals defined by the intent in both situations.

%, leads to better compliance and consistency in Quality of Service (QoS) compared to less efficient methods. This interaction allows ZTN to assure IBN of the expected throughput's compliance with established standards. Utilizing declarative intents and ML methods, this framework offers an extensible solution for intelligent network system design.

\end{abstract}
% \begin{keywords}

\section{Introduction}

As we move towards the 6G era \cite{khan20206g}, the research attention shifts from simple connectivity to connected intelligence. Future networks are expected to meet demanding requirements, including high reliability, mobility, ultra-fast speeds, low latency, massive connectivity, and seamless integration across diverse applications \cite{wei2020intent}. Achieving these ambitious goals necessitate significant enhancements in network capabilities, driven by increased automation. The vision of intelligent autonomous networks aims to minimize human intervention through features, such as, self-configuration, self-learning, and self-optimization \cite{chergui2022toward}\cite{coronado2022zero}. Such systems can detect issues, dynamically allocate resources, and adapt in real time to ensure continuous operation. The ZTN concept is designed to create an automated and intelligent infrastructure with minimal human oversight for design, monitoring, and optimization. This aligns with standards such as the ETSI Zero-Touch Service Management (ZSM) framework \cite{etsi2019zero}.

% ZTN leverages technologies like programmatic control loops, virtualization, orchestration, data analytics, and AI/ML to support effective decision-making, meet 6G performance requirements, and mitigate operational issues \cite{cerda2023anomaly}.
% The stringent demands of 6G applications, such as efficient resource allocation, spectrum management, and rapid response, require intelligence and automation. However, networks must also address challenges like congestion, security threats, and link failures.

Despite the growing adoption of AI and ML in network automation, deploying ML models still presents challenges, particularly in terms of scalability and real-time decision-making. Advancements in network slicing and resource management, along with tools like AutoML \cite{yang2024enabling}, have simplified ML model development for ZTN security.
%However, a fully automated closed-loop control system for end-to-end network optimization remains to be explored in depth.
Adaptive network management requires a resilient closed-loop framework with intelligent analysis, automated actions, and real-time monitoring \cite{bhattacharya2023zero}. Future networks must autonomously fulfill user intent, shifting from manual configurations to IBN. It expresses goals through Graphical User Interfaces (GUI), templates, Natural Language Processing (NLP) \cite{el2023intent}, or intent languages like YANG, LAI, and NEMO \cite{hadi2023survey}, making intent expression more accessible, especially for non-technical users than traditional configurations. It interprets user intents, translates them into actionable policies and enforces user-specific configurations to ZTN. Then, ZTN needs to continuously monitor and act on the network to ensure alignment with these intent based polices which lead seamless and autonomous networks.

\begin{figure}[htbp!]
    \centering
     \includegraphics[width=0.8\columnwidth]{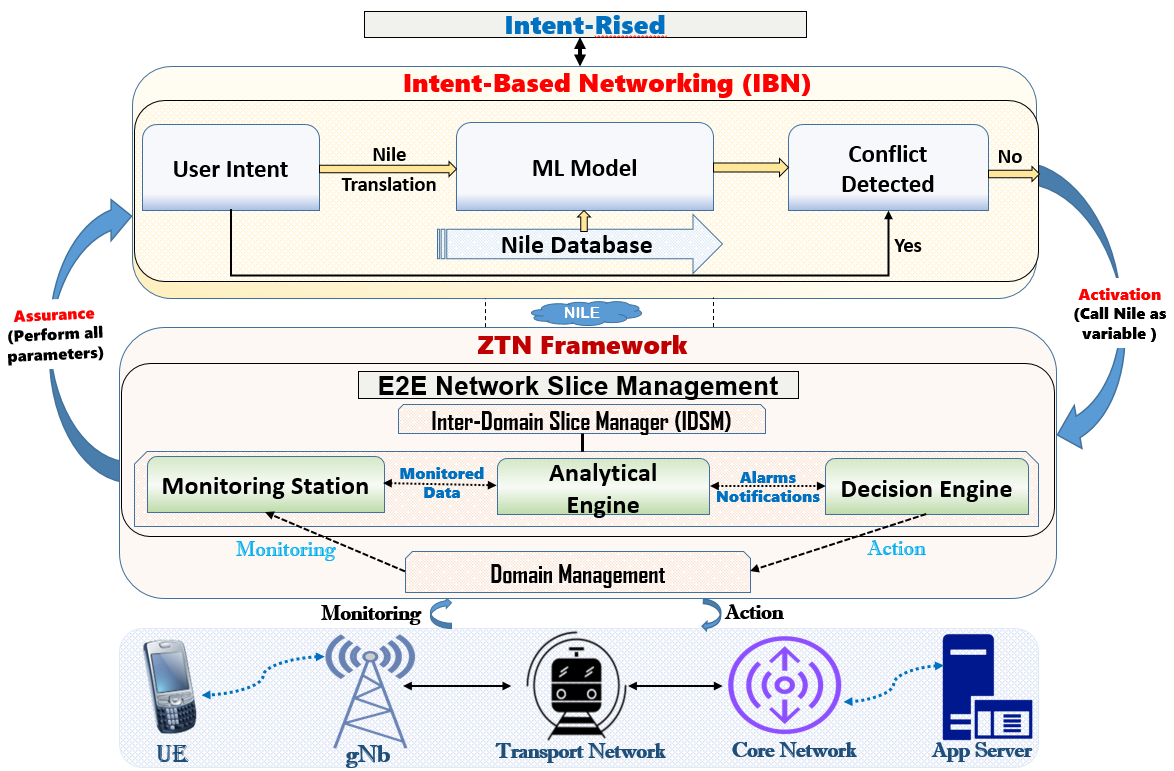}
    \caption{Integrated architecture of  ZTN  and IBN implemented on \textit{OpenAirInterface}}
    \label{Integration IBN-ZTN Framework}
\end{figure}

\begin{figure}
\centering
\includegraphics[width=0.8\columnwidth,height=4cm]{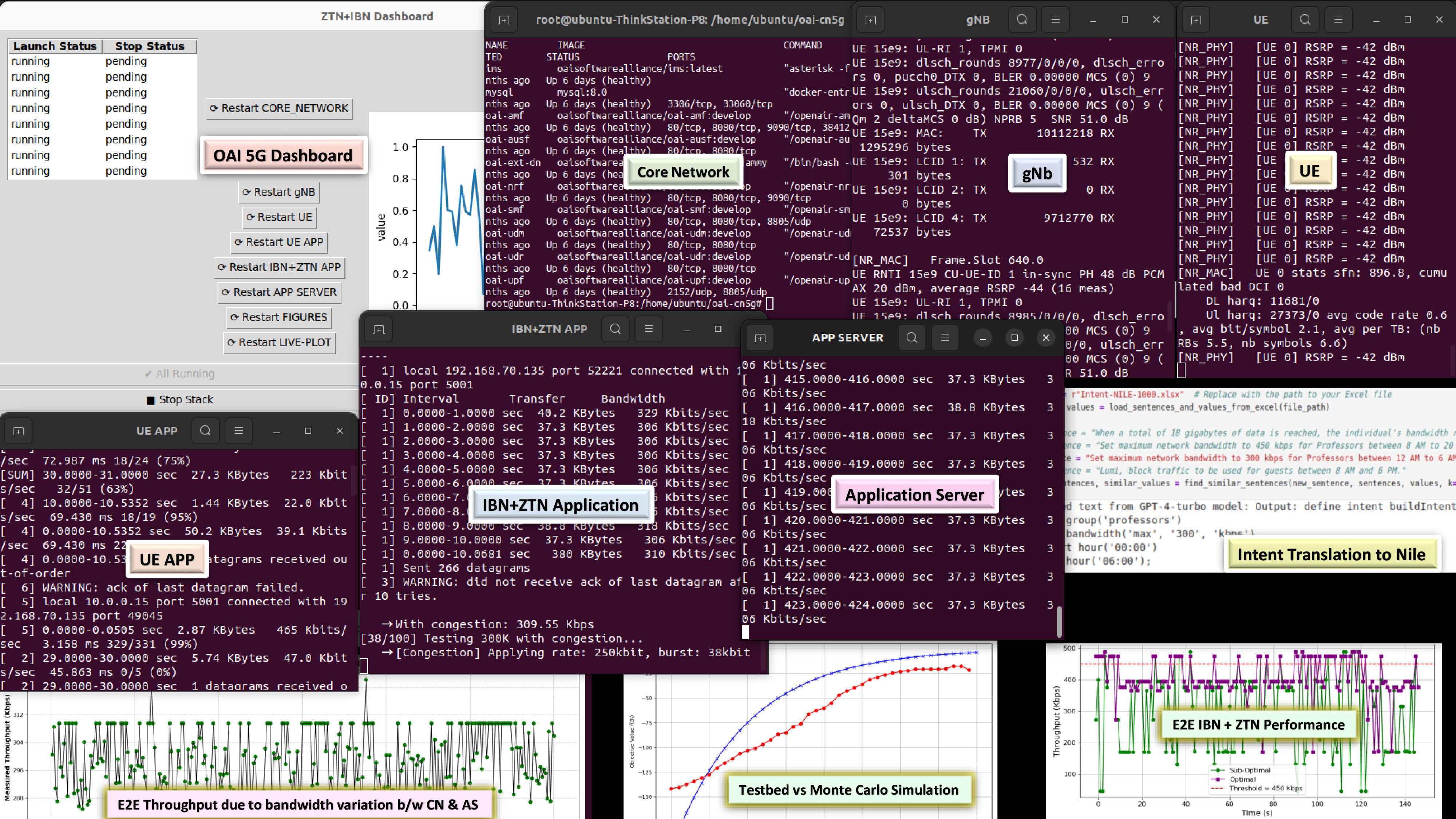}
\caption{Screenshot of the OAI testbed}
\label{fig_oai_screenshot}
\end{figure}

With this motivation, this work proposes an architecture to integrate IBN and ZTN. The intention is that once an intent is set by the user in the IBN framework the ZTN can ensure that the intent is served continuously and autonomously.
%It uses deep learning-powered predictions and bandwidth thresholds based on user intent.
The proposed integrated of ZTN framework and IBN is shown in Fig. \ref{Integration IBN-ZTN Framework}. This architecture is also implemented in OAI platform (Fig. \ref{fig_oai_screenshot}).
The user intent is provided in natural language, e.g., English. The IBN block converts it into intermediate Nile language using a retrieval augmented generation (RAG) based approach \cite{natarajan2025rag}. The IBN component also does basic conflict detection of user intents. The Nile intent is forwarded to the ZTN framework. The ZTN framework with its BiLSTM based network state prediction and a Q-learning based closed loop control maintains the user intent requirements under varying network conditions \cite{tamizhelakkiya2025novel}. Thus, the proposed architecture can autonomously serve, with the user just specifying its usage requirement in natural language English. Rest of the assurance of the intent is taken care by the AI/ML models in the ZTN network.

Two types of scenarios are considered to evaluate the proposed architecture. Firstly, an in-distribution (ID) intent scenario is considered where the ZTN framework has already learnt the actions to maintain a certain range of intent QoS requirements (e.g. bandwidth). Secondly, an out-of-distribution (OOD) case is experimented to evaluate the performance of the ZTN framework with a new intent QoS requirement unknown to it. These two scenarios are measured to find out if the expected values of QoS parameters (e.g., bandwidth) match the user intent requirements over a period of time under varying network conditions. Experimental results show how the proposed architecture is able to handle requirement for end-to-end (E2E) throughput to serve the user intents satisfactorily. To the best of the knowledge of the authors, no prior work has proposed such an integrated architecture, implemented on a testbed, and moving towards futuristic autonomous networks.

%Using LSTM-based predictions over multiple episodes, we evaluate ZTN's throughput performance and define time-sensitive bandwidth intents for various user groups using the Nile language. These goals help us in establishing precise bandwidth expectations, such as 250 Kbps (kilobits per second) for in-distribution (ID) scenarios and 450 Kbps for out-of-distribution (OOD) scenarios, which we subsequently dynamically examine and utilize as assessment benchmarks. We evaluate if the expected values match or exceed the bandwidth restrictions specified by Nile by averaging throughput measurements, which are grouped every five time steps. Plot graphs that contrast the LSTM-predicted throughput for each intent condition for both episodes are used to illustrate this data. This method shows how we can convert IBN intents into real-time performance measurements within ZTN systems, in addition to highlighting the agent's learning behavior.

%Our approach connects operational automation (ZTN) with policy development (IBN), creating a feedback loop that allows continuous monitoring of network activity against the objectives specified by network administrators. This is crucial for building networks that can adapt and validate themselves in response to changing circumstances.

The following are the main contributions of our work: 1) We proposed a novel integrated architecture that combines ZTN with IBN, extracting bandwidth requirement from English language user intents, and programming them into ZTN closed loop as a goal. 2) This architecture is implemented on OAI based testbed (Fig. \ref{fig_oai_screenshot}). 3) Bandwidth variations are introduced in the network and ZTN closed loop is trained to handle these changing conditions to meet the E2E throughput requirements of user intents perpetually using traffic shaping (TS) actions.
%2) We develop a mechanism for throughput analysis to calculate the average bandwidth at predetermined intervals, allowing for a detailed assessment of system performance over time. 3) We implemented a threshold matching technique to determine the frequency with which successful actions meet the predetermined bandwidth targets throughout each episode, ensuring consistency and dependability in action selection.
4) We formulate an optimization problem to evaluate the proposed integrated architecture. The optimization problem is evaluated with Monte Carlo simulations.
5) MOS values are also measured for QoE to indicate the user satisfaction.
6) We present a comparative analysis of the effectiveness of optimal and suboptimal ZTN actions in both ID and OOD user intent scenarios. Results show how the ZTN can ensure that user intents satisfactorily. %Describes how policy alignment and the quality of actions taken affect the function of the system.

%\begin{itemize}
%  \item We developed a novel framework that combines ZTN with intent-based bandwidth rules, extracting bandwidth restrictions from high-level IBN specifications and programming them as variables.
%  \item We developed a pipeline for throughput analysis to calculate the average expected bandwidth at predetermined intervals, allowing for a detailed assessment of system performance over time.
%  \item We implemented a threshold matching technique to determine the frequency with which successful actions meet the predetermined bandwidth targets throughout each episode, ensuring consistency and dependability in action selection.
%  \item We presented a comparative analysis technique to evaluate the effectiveness of optimal and suboptimal ZTN actions in both ID and OOD scenarios. Describes how policy alignment and the quality of actions taken affect the function of the system.

%\end{itemize}

The rest of the paper is structured as follows. We describe our proposed novel integrated architecture in Section II. The results and discussion of our experiments are presented in Section III. Section IV concludes the paper conclusions with directions for future research.

\vspace{-0.2cm}
\section{System Model}\label{section_system_model}
%This section explains how the system uses a ZTN closed-loop pipeline to handle network layout, IBN, and ZTN for making decisions based on bandwidth, as illustrated in Fig. 1. It evaluates bandwidth predictions and action effectiveness in real time using four key parameters: extracting bandwidth requirements from IBN intents, computing average throughput across time slots, counting threshold exceedances, and comparing suboptimal versus optimal actions—enabling adaptation to dynamic traffic patterns.
%This section describes how the system employs a ZTN closed-loop pipeline to manage network topology, IBN, and ZTN for making decisions based on bandwidth, as shown in Fig. 1. It assesses bandwidth predictions and effectiveness of action in real time based on four crucial parameters: extracting bandwidth needs from IBN intents, calculating average throughput over time slots, threshold exceedance counting, and comparing suboptimal and optimal actions—facilitating adaptation to changing traffic patterns. This integration involves nine key phases that outline the system's architecture. Detailed explanation of each component is provided below.

This section describes the proposed architecture of integrating IBN and ZTN. It involves translation of user intent specified in natural language (English) to Nile syntax. This Nile intent is then passed on to the ZTN subsystem as a goal. ZTN closed loop job is to assure that the intent is sustained under varying network condition (e.g. bandwidth variations).

\vspace{-0.2cm}
\subsection{Translation of Intent}
%TBD: This section can be reduced as this are already done in Uma's paper. reference to Uma' paper should be enough.
This function aims to convert natural language intents into executable Nile code using large language models (LLMs) using the methodology  explained in \cite{natarajan2025rag}.
This approach performs efficient  translation of user intents in IBN context using the Retrieval-Augmented Generation (RAG) inspired approach \cite{lin2023appleseed}. The method translates an intent in English to Nile format using a few-shot prompting approach with zero training and a small dataset. It also has a conflict detection mechanism with previously configured intents. The user-provided natural language intent $I_{\text{NL}}$ is translated into the NILE format $I_{\text{NILE}} = \beta_{\text{target}} \in \mathbb{R}^{+}$ using a Retrieval-Augmented Generation (RAG)-based LLM translator $T_{\text{LLM}}$, in equation 1.
\begin{equation}
I_{\text{Nile}} = T_{\text{LLM}}(I_{\text{NL}}) \quad \text{where} \quad I_{\text{Nile}} = \beta_{\text{target}} \in \mathbb{R}^{+}
\end{equation}

\noindent
\text{Where:}
\begin{itemize}
    \item $I_{\text{NL}}$ : User intent in Natural Language
    \item $I_{\text{Nile}}$ : Intent translated into NILE format (e.g., bandwidth requirement $\beta_{\text{target}}$)
    % \item $T_{\text{LLM}} : I_{\text{NL}} \rightarrow I_{\text{Nile}}$ : RAG-based intent translator
    \item $T_{\text{LLM}} (.)$:  RAG-based intent translator
\end{itemize}

\vspace{-0.2cm}
\subsection{Network Topology}
Network topology comprises a core network (CN) connected with application servers. gNBs are connected with CN and support user equipments (UEs). Variations QoS parameters, i.e., bandwidth, latency, jitter, and packet loss, may be experienced during data communications between UEs and the application servers. These variations of parameters may be saved as a dataset for training the ZTN closed loop control.

\vspace{-0.2cm}
\subsection{ZTN Closed Loop}
The BiLSTM model is trained on network dataset to predict network states, such as, bandwidth, for proactive decisions and resource optimization as explained in \cite{tamizhelakkiya2025novel} %[TBD: Elakkiya's name is not there in the bibtex. Never use bibtex citation from google scholar, always from the journal or archive website]. %It processes inputs in both forward and backward directions to improve accuracy.
The predicted bandwidth at time $t+1$, denoted as $\hat{B}_{t+1}$, is generated using a BiLSTM model $f_{\text{BiLSTM}}$ applied over the past $n$ time steps of bandwidth $B_{t-n:t}$, with model parameters $\theta$, in equation (2).
Predictions of the BiLSTM model are compared with the ground truths to compute residuals, which are then used to train an XGBoost model. This secondary model learns and corrects the BiLSTM's errors. The resulting hybrid model enhances network state prediction accuracy. Given the predicted state $S_t = \hat{B}_{t+1}$, the Q-learning agent selects an optimal action $A_t$ using a policy $\pi(S_t) = \arg\max_{a \in \mathcal{A}} Q(S_t, a)$, and updates the Q-value using the Bellman equation, as shown in equations (3) and (4).

\begin{equation}
\hat{B}_{t+1} = f_{\text{BiLSTM}}(B_{t-n:t}; \theta)
\end{equation}

% The predicted bandwidth at time $t+1$, denoted as $\hat{B}_{t+1}$, is generated using a BiLSTM model $f_{\text{BiLSTM}}$ applied over the past $n$ time steps of bandwidth $B_{t-n:t}$, with model parameters $\theta$.
% Let $B_t$ denote the actual bandwidth at time $t$. The predicted bandwidth at time $t+1$, denoted as $\hat{B}_{t+1}$, is computed using a BiLSTM model as:
where $B_{t-n:t}$ represents the sequence of bandwidth values from time $t-n$ to $t$, and $\theta$ denotes the learned parameters of the BiLSTM model.

\begin{equation}
A_t = \pi(S_t) = \arg\max_{a \in \mathcal{A}} Q(S_t, a)
\end{equation}

\begin{equation}
\small
Q(S_t, A_t) \leftarrow Q(S_t, A_t) + \alpha \left[ R_t + \gamma \max_{a'} Q(S_{t+1}, a') - Q(S_t, A_t) \right]
\end{equation}

% Let $S_t = \hat{B}_{t+1}$ represent the predicted network state at time $t$, where $\hat{B}_{t+1}$ is obtained from the BiLSTM model.

% The action selected by the Q-learning agent at time $t$ is denoted by $A_t \in \mathcal{A}$, where $\mathcal{A}$ is the set of possible actions (e.g., traffic shaping or delay control).

% The policy $\pi$ selects the optimal action based on the predicted state as:
% \begin{equation}
% \pi(S_t) = \arg\max_{a \in \mathcal{A}} Q(S_t, a)
% \end{equation}
where $Q(S_t, a)$ is the Q-value representing the expected future reward for taking action $a_t$ in state $S_t$. Action $A_t$ will allocate bandwidth $B_t^{(a)}$.

The reward at time $t$ is denoted as $R_t$, which is computed based on how well the selected action meets the intent defined by $I_{\text{Nile}}$.
% Given the predicted state $S_t = \hat{B}_{t+1}$, the Q-learning agent selects an optimal action $A_t$ using a policy $\pi(S_t) = \arg\max_{a \in \mathcal{A}} Q(S_t, a)$, and updates the Q-value using the Bellman equation.

%The BiLSTM model \cite{el2023intent} is trained on network topology data to predict network states, including bandwidth estimation, enabling proactive decision-making, optimized resource allocation, and congestion avoidance. It is trained in two directions: the forward path processes initial input samples in sequence, while the backward path uses BiLSTM layers on an inverted version of the input to enhance prediction accuracy. During testing, input samples are fed into the trained BiLSTM model to predict network states. The residuals—differences between predicted and actual values—are then used to train an XGBoost model \cite{el2023intent}, which learns the error patterns of the BiLSTM predictions. This secondary model refines the initial predictions by correcting residual errors, producing its own residuals after training. The resulting hybrid model \cite{das2025novel}, combining BiLSTM and XGBoost, predicts QoS parameters more accurately, with performance evaluated using the mean squared error (MSE) metric.
%\vspace{-0.2cm}
%\subsection{Optimal Action Selection}
The hybrid model and  Q-learning model try to select actions based on predicted network states \cite{zhuang2022short}. Each state reflects network conditions, allowing ZTN to select optimal configurations as actions and proactively enhance performance perpetually in a closed loop.

\vspace{-0.2cm}
\subsection{Extraction of Bandwidth from Nile Intent}
The Nile intent acts as a guiding policy for the ZTN framework to function. For bandwidth requirements, using the specified clause: $set bandwidth$ ${'max',B_i,'kbps'}$ where $B_i$ indicates the expected bandwidth of the user intent. In this approach, we extract the value of $B_i$ from the Nile intent using regular expressions and is then set as a goal for the ZTN system. We considered two different goal thresholds to see how the system performs. The ID scenario reflects expected or previously observed traffic patterns, and the OOD scenario represents unexpected or new network goals. %These thresholds are crucial in determining whether the ZTN system can meet the QoS defined by the intent, ensuring it can make real-time decisions that align with high-level IBN policies.

\vspace{-0.2cm}
\subsection{Evaluation of Suboptimal Vs Optimal Action}
To measure the performance of Q-learning, we evaluate and contrast system performance in two different scenarios. The optimal action case, in which decisions are trained into the Q-learning model, and the sub-optimal condition, in which the ZTN system makes judgments without being completely trained.

% The reward $R_t$ is assigned based on whether the actual bandwidth $B_{t+1}$ meets or exceeds the target intent $\beta_{\text{target}}$, where $R_t = +1$ if $B_{t+1} \geq \beta_{\text{target}}$, and $R_t = -1$ otherwise.

% For each episode $e \in \mathcal{E}$, the objective of the closed-loop system is to maximize the number of time steps $\delta_t$ for which the bandwidth $B_t$ satisfies the intent, i.e., $\sum_{t=1}^{T} \delta_t \rightarrow \max$.

\begin{equation}
R_t =
\begin{cases}
+1, & \text{if } B_{t+1} \geq \beta_{\text{target}} \\
-1, & \text{otherwise}
\end{cases}
\end{equation}

The reward $R_t$ is assigned based on whether the actual bandwidth $B_{t+1}$ meets or exceeds the target intent $\beta_{\text{target}}$, where $R_t = +1$ if $B_{t+1} \geq \beta_{\text{target}}$, and $R_t = -1$ otherwise, in equation (5).
Let $\mathcal{E}$ denote the set of network episodes over which the model operates. For each episode $e \in \mathcal{E}$, the objective of the closed-loop system is to maximize the number of time steps $\delta_t$ for which the bandwidth $B_t$ satisfies the intent over a period of time $T$
%, that is, $\sum_{t=1}^{T} \delta_t \rightarrow \max$,
in equations (6) and (7).

\begin{equation}
\forall e \in \mathcal{E}, \quad max \sum_{t=1}^{T} \delta_t %\rightarrow \max
\end{equation}

\begin{equation}
\delta_t =
\begin{cases}
1, & \text{if } B_t \geq \beta_{\text{target}} \\
0, & \text{otherwise}
\end{cases}
\end{equation}

Here, $\delta_t = 1$ indicates that the actual bandwidth at time $t$ meets or exceeds the target intent threshold $\beta_{\text{target}}$.

The optimal action $A_t$ is selected by a Q-learning policy based on the BiLSTM-predicted network state, ensuring that the resulting bandwidth $B_{t+1}$ meets the threshold derived from the user's natural language intent via a RAG-based LLM translator, in equation (8).

\begin{equation}
A_t = \pi \left( f_{\text{BiLSTM}}(B_{t-n:t}; \theta) \right), \quad \text{subject to: } B_{t+1} \geq T_{\text{LLM}}(I_{\text{NL}})
\end{equation}

\noindent
\text{Where:}
\begin{itemize}
    \item $A_t$: Action selected by the Q-learning policy at time $t$
    \item $f_{\text{BiLSTM}}(B_{t-n:t}; \theta)$: BiLSTM model predicting future bandwidth based on previous $n$ time steps
    \item $\pi(\cdot)$: Q-learning policy that selects the optimal action based on the predicted state
    \item $B_{t+1}$: Actual bandwidth observed after applying action
    \item $T_{\text{LLM}}(I_{\text{NL}})$: Bandwidth threshold is determined using a RAG-based LLM to interpret the user's natural language intent.
\end{itemize}

The total user experience is measured by the expected Mean Opinion Score (MOS), which measures Quality of Experience (QoE). This score is based on how often the network successfully reaches the user's intended bandwidth threshold. This score is a human-centered measure of network performance. It is the normalized ratio of time steps where the actual bandwidth meets the IBN-derived target, scaled between a minimum and maximum MOS range, in equation (9).

%Let $\delta_t = 1\{ B_t \geq \beta_{\text{target}} \}$ be the indicator of bandwidth satisfaction at time $t$, where $B_t$ is the actual bandwidth and $\beta_{\text{target}}$ is the threshold from the IBN-derived intent.

The predicted MOS \cite{itutp800} score is computed as:

\begin{equation}
\text{MOS}_{\text{pred}} = \text{MOS}_{\min} + \left( \frac{1}{T} \sum_{t=1}^{T} \delta_t \right) \cdot (\text{MOS}_{\max} - \text{MOS}_{\min})
\end{equation}

\noindent
Where:
\begin{itemize}
    \item $T$: Total number of time steps
    \item $\text{MOS}_{\max}$: Maximum MOS score (ID: 5 for excellent QoE)
    \item $\text{MOS}_{\min}$: Minimum MOS score (OOD: 1 for poor QoE)
\end{itemize}

\subsection{Optimization formulation}
The following optimization is formulated.

Objective:
\begin{align}
% O_1: \quad & \max_{a_t \in A} \left( \beta_{\text{target}} - B_t^{(a)} \right) \\
O_1: \quad & \min_{a_t \in A} | \beta_{\text{target}} - B_t^{(a_t)}|\\
% O_2: \quad & \max_{a_t} \; B_t^{(a)} \rightarrow B_{t+1} \\
%O_2: \quad & \max_{a_t \in A} \text{MOS}_{\text{pred}}
% = \text{MOS}_{\min} + \left( \frac{1}{T} \sum_{t=1}^{T} \delta_t \right) \cdot \left( \text{MOS}_{\max} - \text{MOS}_{\min} \right)
\end{align}

\text{subject to,} \notag \\
% \[c_1: \quad & B_{\min} \leq B_t^{(a)} \leq B_{\max} \notag \]
\begin{align}
c_1: \quad B_{\min} \leq B_t^{(a_t)} \leq B_{\max} \notag
\end{align}
where $B_{min}$ and $B_{max}$ denotes the lower and upper bounds of bandwidth provided by the underlying network.

% a_t = {\quad & \hat{B}_t \notin D_{\text{train}} \Rightarrow a_t = a_{\text{fallback}}
% \end{align}

\[c_2:
a_t =
\begin{cases}
a_{\text{optimal}}, & \text{if } \hat{B}_t \in D_{\text{train}} \\
a_{\text{suboptimal}}, & \text{if } \hat{B}_t \notin D_{\text{train}}
\end{cases}
\]
where $D_{\text{train}}$ is the set for ID, and $a_{optimal}$ and $a_{suboptimal}$ are actions chosen during ID and OOD scenarios.

The above optimization problem is evaluated using Monte Carlo simulations.
\vspace{-0.1cm}
\section{Results and Discussion}\label{resul_discu}
This section presents the results from the testbed based on our system model. We will also show how the closed-loop control mechanism works for ZTN and IBN integration based on BiLSTM-based bandwidth prediction. This integration is for dynamic QoS assurance of user intents. For simplicity, we are considering only the downlink E2E bandwidth as our QoS parameter. %By extracting bandwidth thresholds from the IBN intents described in Nile and monitoring ZTN's performance through time-grouped throughput logs, we've created a closed-loop system.
This allows ZTN to constantly check whether its predicted throughput meets or exceeds the intent-driven thresholds. Naturally, other parameters can also be taken into account in examining our generalized system model. We use an \textit{i9} CPU and 64 GB RAM based system for this research with OAI. The models are all implemented in python programming language.

\begin{figure}
\centering
\includegraphics[width=0.8\columnwidth,height=4cm]{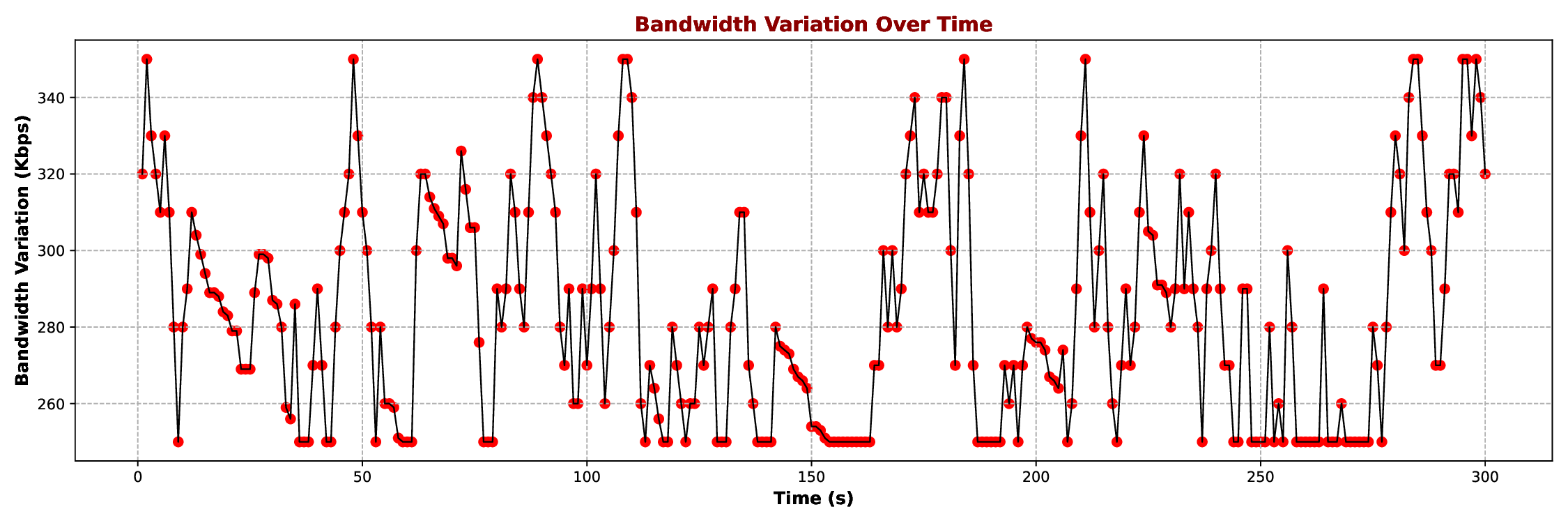}
\caption{Injected bandwidth variation between CN and AS}
\label{fig_bandwidth_variation_upf_as}
\end{figure}

\begin{figure}
\centering
\includegraphics[width=0.8\columnwidth,height=4cm]{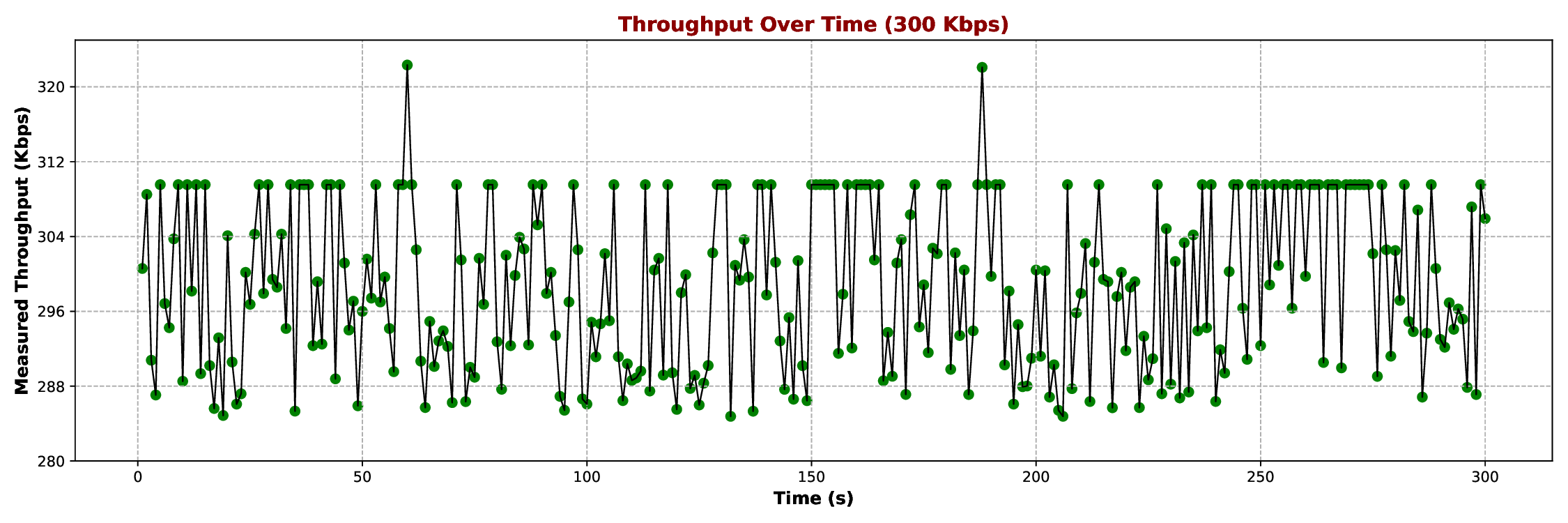}
\caption{E2E throughput due to bandwidth variation between CN and AS without ZTN closed loop control (ID scenario)}
\label{fig_throughput_plot_300kbps}
\end{figure}

\begin{figure}
\centering
\includegraphics[width=0.8\columnwidth,height=4cm]{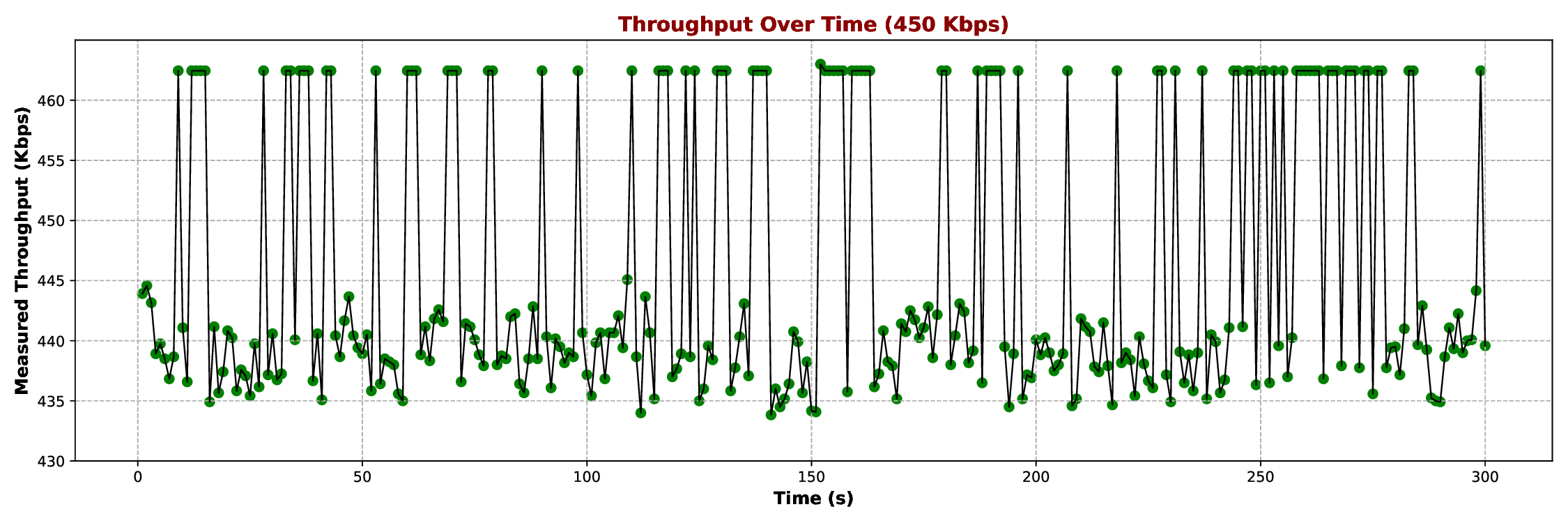}
\caption{E2E throughput due to bandwidth variation between CN and AS without ZTN closed loop control (OOD scenario)}
\label{fig_throughput_plot_450kbps}
\end{figure}
\vspace{-0.2cm}
\subsection{Experimental Setup}
The E2E network topology uses the OAI platform which includes a Core Network (CN) with gNB, User Plane Function (UPF), Session Management Function (SMF), and Access and Mobility Management Function (AMF), enabling uplink/downlink data flow from UE through the UPF to the AS. The RAN comprises gNB and UEs. %, with UE arrivals following a Poisson distribution. UE join rates are mapped to the observed bandwidth to evaluate ZTN performance.
The downlink E2E throughput between AS and UE, recorded over 100 seconds when bandwidth variations are applied to the interface between UPF and the AS. The collected time series data are used to train the BiLSTM model offline. The Q-learning training with TS happens online.

We used \textit{iperf} for controlled testing within an OAI 5G testbed environment to determine the influence of the injected network bandwidth variation. Bandwidth variations is simulated randomly for 300 randomized bandwidth values. Traffic control (\textit{tc}) command was utilized at the UPF interface to simulate bandwidth variation by enforcing burst sizes and variable rate bounds. The E2E throughput was observed for bandwidth variations at the UPF, and these results are shown in Fig. \ref{fig_bandwidth_variation_upf_as}. Figs. \ref{fig_throughput_plot_300kbps} and \ref{fig_throughput_plot_450kbps} respectively show the E2E throughputs for ID and OOD scenarios of 300 Kbps and 450 kbps user intents without the ZTN functionality. These are raw throughputs for the two scenarios. As will be seen latter, applying ZTN closed loop actions leads to sustained throughput as required by the respective intents. %,  these plots graphically depict performance loss with variation. In a simulated 5G core network, the collected data and corresponding graphs clearly illustrate a correlation between the imposed congestion rates and reduced performance, as shown in Fig. 3 (b) for ID scenario and (c) for OOD scenario, affirming the responsiveness and resilience of the end-to-end performance system, underscoring the importance of adaptive traffic management in real implementations.
\subsection{Intent Translation}
\begin{figure}[htbp!]
    \centering
    \includegraphics[width=0.8\columnwidth,height=2cm]{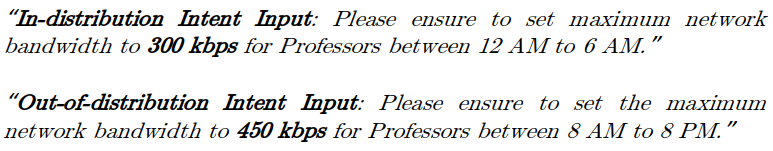}
    \caption{In-Distribution and Out-of-Distribution Input Intents}
    % \textit{OpenAirInterface}}
    \label{fig_Input_Intents}
\end{figure}

\begin{figure}[ht!]
    \centering
    \includegraphics[width=0.8\columnwidth,height=4cm]{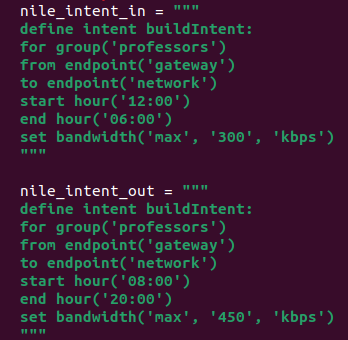}
    \caption{Nile Intent for in-distribution and out-of-distribution}
    \label{fig_Nile_Intent}
\end{figure}

ID and OOD User intents in English show in Fig. \ref{fig_Input_Intents} are provided to the RAG based translator. The RAG based model translates this to Nile format (Fig. \ref{fig_Nile_Intent}). This Nile intent is then passed on to the ZTN framework as a goal to be ensured by its closed loop.

\vspace{-0.2cm}
\subsection{ZTN Closed Loop Performance}
For predicting the network state, i.e., bandwidth, E2E throughput for ID scenario is used for training the BiLSTM model. During inference on live downlink data transmission, the predicted state is passed on the Q-learning based closed loop to choose the appropriate traffic shaping action so that the user intent is met.

\begin{figure}
\centering
\includegraphics[width=0.8\columnwidth]{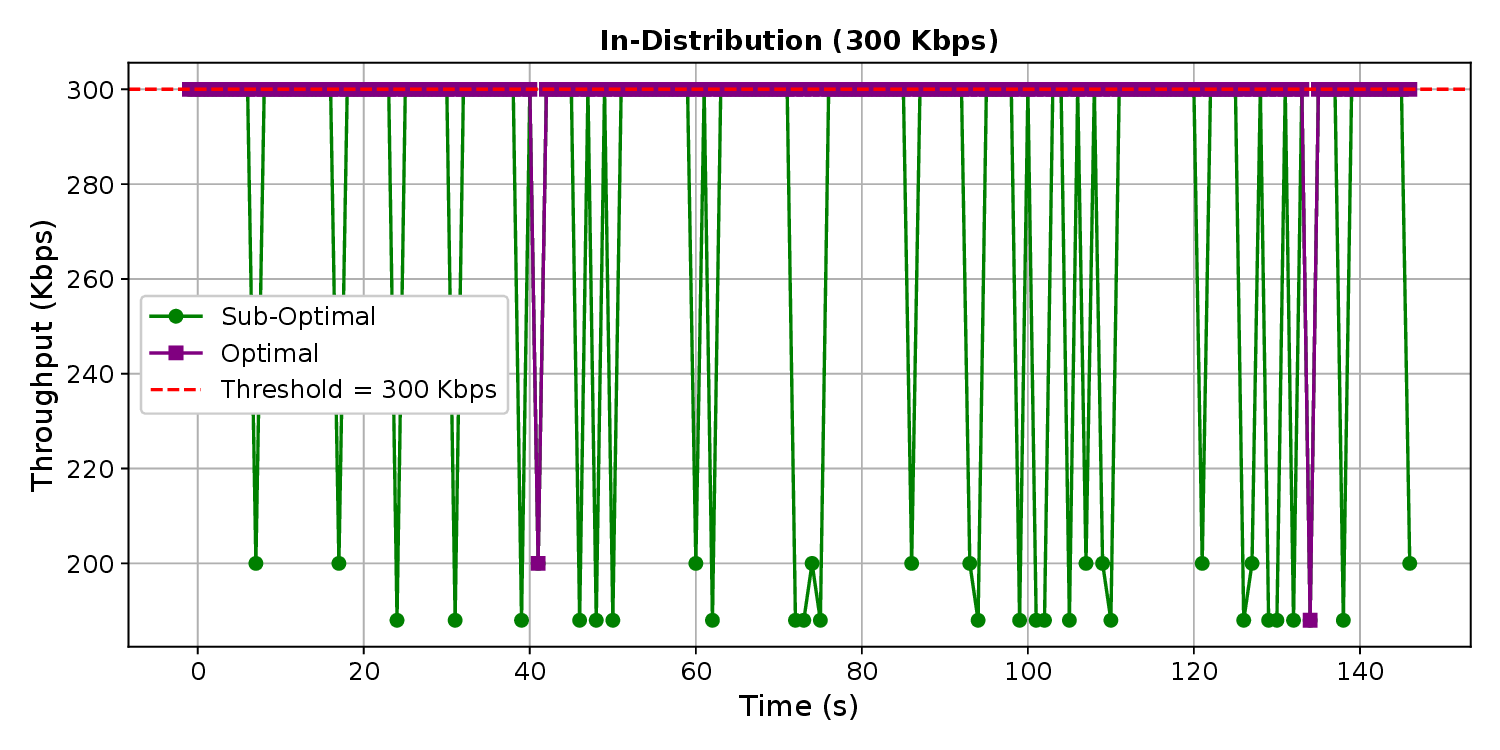}
\caption{E2E throughput achieved applying ZTN closed loop to meet ID user intent of 300 kbps}
\label{fig_e2e_througput_id}
\end{figure}

In the ID case, we assign the E2E throughput  goal of 300 kbps from the Nile intent. %In this case, optimal and sub-optimal actions are likely to closely match the intent-determined requirement.
%This threshold is used as a benchmark to check whether the predicted E2E throughput is meeting the user intent.
As indicated in Fig. \ref{fig_e2e_througput_id} for the optimal scenario when the Q-learning is fully trained it meets the intended throughput of 300 kbps most of the time. Interesting, even for sub-optimal case (when the Q-learning model has not fully learnt), there are several occasions the intent threshold is met.

\begin{figure}
\centering
\includegraphics[width=0.8\columnwidth]{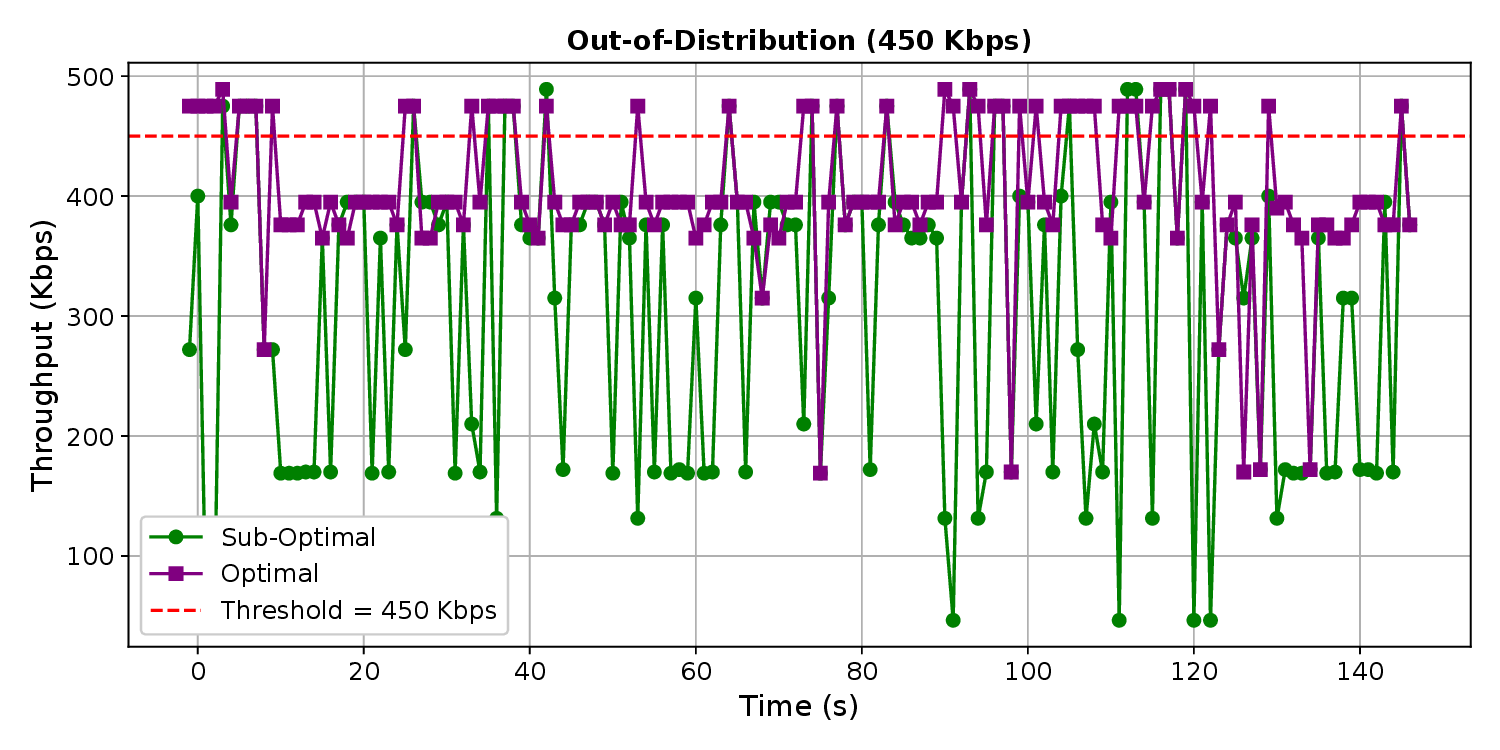}
\caption{E2E throughput achieved applying ZTN closed loop to meet OOD user intent of 450 kbps}
\label{fig_e2e_througput_ood}
\end{figure}

In the OOD case of 450 kbps threshold in Fig. \ref{fig_e2e_througput_ood}, when the Q-learning model has to ensure a new goal, even for the optimal scenario there are several instances when the intent is not met. This is because the current actions known to the ZTN closed loop are not sufficient to meet the intent requirements. Hence, in this case, more exploration of the action space is necessary. %as the current set of actions does not meet the intent requirements.
As expected, in the suboptimal case the performance is also in similar lines with lower performance.

The predicted ($\text{MOS}_{\text{pred}}$) are computed by taking a time-normalized average of the number of time steps in which actual throughput meets or exceeds the IBN defined intent bandwidth goal.
% We use a scale that is bounded by $\text{MOS}_{\min} = 1$, in compliance with the ITU-T P.800 guidelines. Here, $\text{MOS}_{\min} = 1$ indicates an exceptionally poor user experience, while $\text{MOS}_{\max} = 5$ indicates an excellent experience.
Out of 148 evaluation runs in the ID scenario, 146 runs met the intent goal ($\delta_t = 1$) with the optimal actions and 112 runs for the suboptimal ones. In contrast, in the OOD scenario, only 47 optimal and 23 suboptimal runs satisfied the same requirement. % \(\beta_{\text{target}}\). %This mapping allows us to convert binary intent satisfaction signals ($\delta_t$) into a continuous human-interpretable QoE metric over time.
The observed MOS values for both ID and OOD situations are summarized in Table 1.
% It has been observed that, in the in-distribution (ID) case, the Mean Opinion Score (MOS) values are 3.8 (Good) for the suboptimal allocation and 4.6 (Excellent) for the optimal allocation. In the out-of-distribution (OOD) case, the observed MOS values are 1.4 (Poor) and 2.2 (Average) for the suboptimal and optimal allocations, respectively.
%We will look the divergenc of OOD by performing a further exploration of the action space in our future work.

\begin{table}[h!]
\scriptsize
% \small
% \begin{table}[h!]
\centering
\caption{Mean Opinion Score Comparison for In-Distribution and Out-of-Distribution Scenarios}
\label{tab:MOS_comparison}
\begin{tabular}{|l|l|c|c|}
\hline
\textbf{Scenario} & \textbf{Allocation Type} & \textbf{MOS Value} & \textbf{MOS Rating} \\
\hline
\multirow{2}{*}{In-Distribution (ID)}
                  & Suboptimal              & 3.8                & Good                \\
                  & Optimal                 & 4.6                & Excellent           \\
\hline
\multirow{2}{*}{Out-of-Distribution (OOD)}
                  & Suboptimal              & 1.4                & Poor                \\
                  & Optimal                 & 2.2                & Average             \\
\hline
\end{tabular}
\end{table}

\begin{figure}
\centering
\includegraphics[width=0.8\columnwidth]{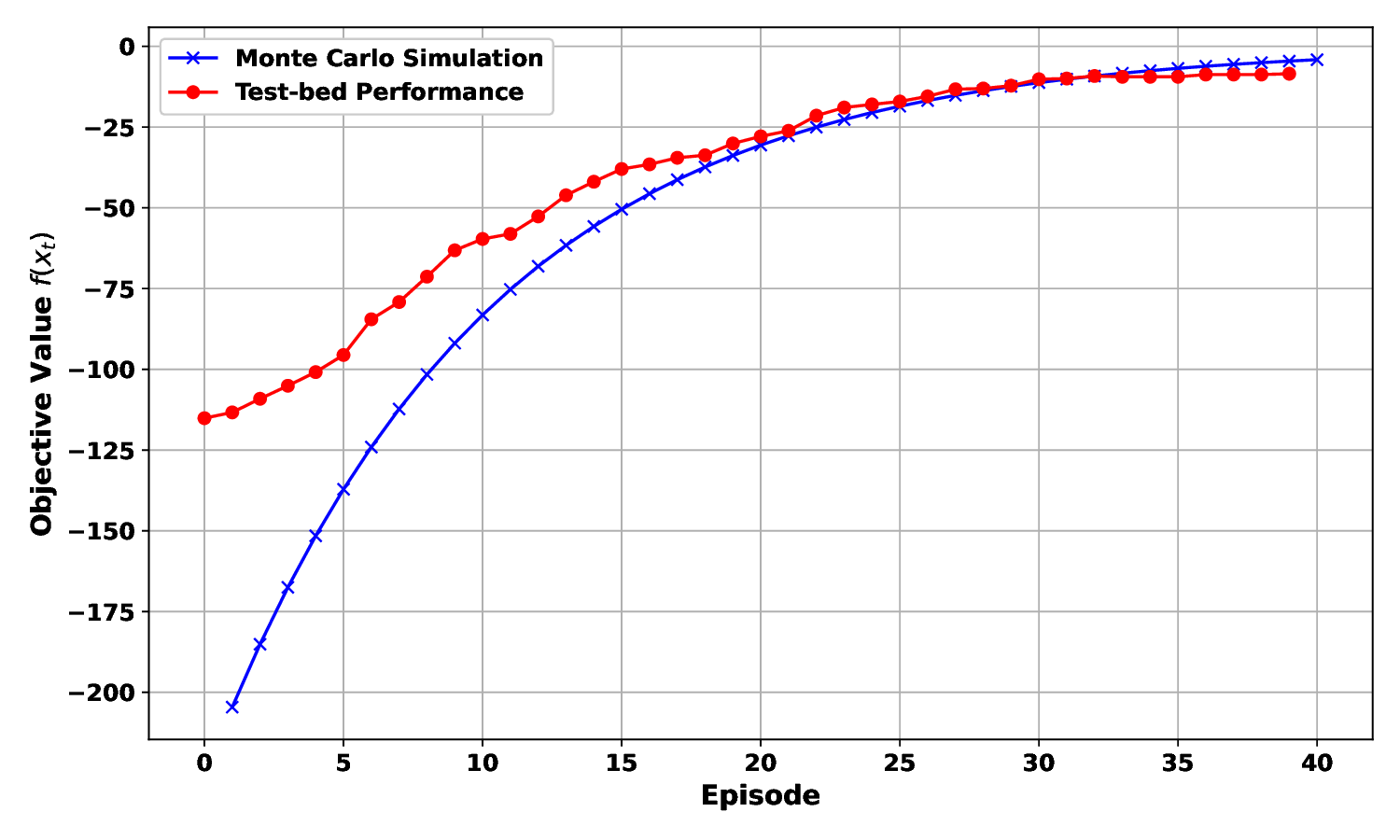}
\caption{Theoretical (Monte Carlo) vs. Practical (Testbed) Performance Under In-Distribution Case}
\label{Monte-carlo-test-bed-compa-ID}
\end{figure}

\begin{figure}
\centering
\includegraphics[width=0.8\columnwidth]{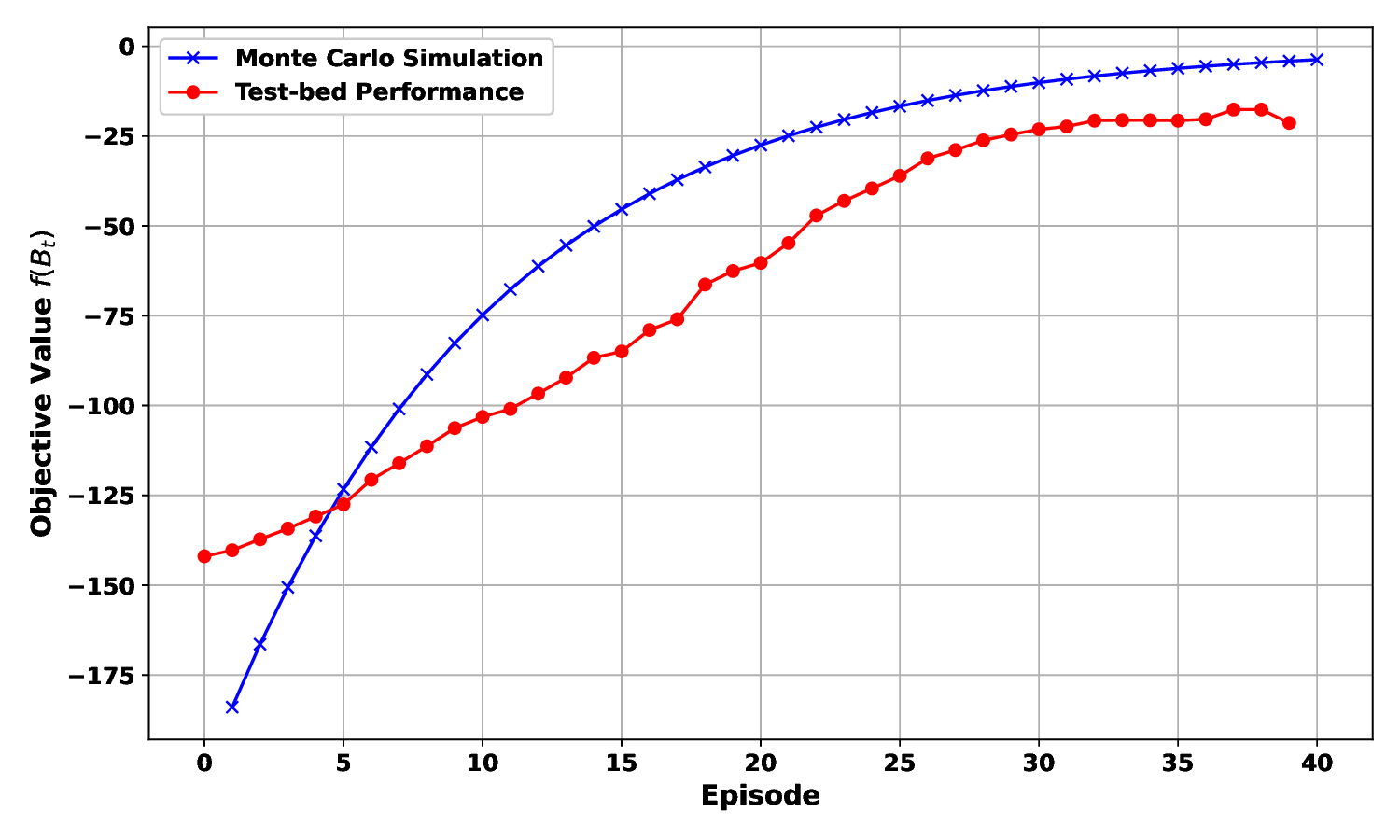}
\caption{Theoretical (Monte Carlo) vs. Practical (Testbed) Performance Under Out-Of-Distribution Case}
\label{Monte-carlo-test-bed-compa-OOD}
\end{figure}

To evaluate the proposed architecture, two observations are compared. Firstly, the formulated optimization problem is evaluated on the testbed. Secondly, Monte Carlo simulation are performed to evaluate the same problem. Both ID and OOD scenarios are considered.
%To evaluate the effectiveness of the proposed system across both in-distribution and out-of-distribution scenarios, the average throughput of each episode is first computed from the testbed results and stored in a CSV file.
The comparison for intent bandwidths - 300 Kbps for ID is shown in Fig. \ref{Monte-carlo-test-bed-compa-ID}, and the same with 450 Kbps for OOD shown in Fig. \ref{Monte-carlo-test-bed-compa-OOD}. %, is then calculated using the expression.
\begin{comment}
\[
\max_{a_t \in A} -\left| \beta_{\text{target}} - B_t^{(a_t)} \right|,
\]
where \( \beta_{\text{target}} \) is the intent-defined bandwidth and \( B_t^{(a_t)} \) is the achieved throughput for action \( a_t \).
\end{comment}
These comparisons are plotted along the Y-axis, with episode indices on the X-axis. Monte Carlo simulation results are shown with blue line while the testbed performance shown in red. In case of ID, the intent goal is reached whereas for the OOD there is a divergence due to sub-optimal actions. However, the trends are similar in both cases. This again highlights the need for perpetual exploration of action space to meet new and unseen intent goals. %This comparative analysis effectively illustrates how closely the practical results align with the theoretical expectations under varying distribution conditions.

\vspace{-0.2cm}
\section{Discussion}
The intents provide a goal that translates to ZTN decisions and actions applied to the underlying network to meet user requests. The BiLSTM model provides network prediction, with optimal actions chosen by ZTN closed loop to closely meet different bandwidth intents under varying network conditions. This analysis confirms that the integration of IBN-specified bandwidth requirements into ZTN control logic supports automated, intent-aware throughput sustenance in a move towards autonomous networks. Though, the ID case performs satisfactorily, perpetual exploration of the action space is required for OOD scenario.

\vspace{-0.2cm}
\section{Conclusion}
Automation is essential for the efficient operation of 6G networks. We proposed a novel architecture that integrates IBN and ZTN. The architecture takes an intent in English, translates into intermediate Nile format using a RAG based approach. The Nile intent is passed on to the ZTN component as a goal to be ensured by the underlying network. ZTN closed loop proactively predicts the network state e.g. bandwidth using XGBoosted BiLSTM model and then Q-learning to provide optimal actions to the network under varying network condition. This architecture is implemented using OAI based testbed. Results show how the architecture can autonomously maintain intents in natural language both for IN and OOD scenarios. Further, an optimization problem is formulated which is evaluated through Monte Carlo simulations and compared with the testbed results which show similar trends. However, for OOD case, there is some amount of divergence and provides scope of further improvement. The MOS scores for user intent QoE also follow similar patterns for ID and OOD scenarios.

Future work will concentrate on considering other QoS parameters beyond bandwidth and exploration of the action space for the OOD scenario.

%By extracting bandwidth thresholds from Nile-defined IBN intents and assessing ZTN’s performance over time-grouped throughput logs, we establish a closed loop system where ZTN actively verifies whether its predicted throughput meets or exceeds the intent-driven thresholds. Experimental results, evaluated in both ID (300 Kbps) and OOD (450 Kbps) scenarios, demonstrate that optimal actions aligned with intents significantly outperform sub-optimal choices. This confirms the practicality of activating intents within ZTN and its role in reinforcing IBN policies in real-time. For future work, we plan to enhance this integration by adding real-time intent updates, prioritizing multiple intents, and orchestrating across different types of networks. We also aim to explore reinforcement learning-based feedback tuning, allowing ZTN to adjust its prediction policies based on feedback from IBN. Combining semantic policy extraction, real-time analytics, and autonomous verification sets the stage for scalable and intelligent networking infrastructures.

\vspace{-0.5cm}
\textcolor{black}{\section*{Acknowledgment}
The authors would like to thank Toshiba Software India Pvt Ltd for sponsoring this research project.}
% This research was supported by the Toshiba Software (India) Private Limited (TSIP) (\url{http://www.toshiba-tsip.com}). The supports are gratefully acknowledged.
\bibliographystyle{IEEEtran}
\bibliography{ZTNbib.bib}

\end{document}